\documentclass[twoside]{article}
\usepackage{fleqn,espcrc2}

\usepackage{graphicx}
\usepackage[figuresright]{rotating}

\newcommand{\AmS}{{\protect\the\textfont2
  A\kern-.1667em\lower.5ex\hbox{M}\kern-.125emS}}

\newcommand{\psibar}{{\overline\psi}}
\newcommand{\ghat}{{\hat\gamma}_5}
\newcommand{\Phat}{{\hat P}}
\newcommand{\vbar}{{\overline v}}
\newcommand{\cbar}{{\overline c}}
\newcommand{\ie}{{\it i.e.}}
\newcommand{\gdofs}{{\it gdofs}}
\newcommand{\eg}{{\it e.g.}}
\newcommand{\Lapl}{\kern0.5pt{\lower0.1pt\vbox{\hrule height.5pt width 6.8pt
    \hbox{\vrule width.5pt height6pt \kern6pt \vrule width.3pt}
    \hrule height.3pt width 6.8pt} }\kern1.5pt}
\newcommand{\etc}{{\it etc}}
\newcommand{\tr}{{\rm tr}}
\newcommand{\Tr}{{\rm Tr}}
\newcommand{\Dsl}{D\!\!\!\!/}
\newcommand{\psl}{p\!\!\!/}
\newcommand{\etal}{{\it et al.}}
\newcommand{\cf}{{\it cf.}}
\newcommand{\tk}{{\tilde\kappa}}

\title{Lattice Chiral Gauge Theories}

\author{Maarten Golterman\vskip 6pt%
{Department of Physics,
        Washington University, 
        St. Louis, 63130 Missouri, USA}}
       
\begin{document}

\begin{abstract}
I review the substantial progress which has been made recently
with the non-perturbative construction of chiral gauge theories 
on the lattice.  In particular, I discuss three different
approaches: a gauge invariant method using fermions which satisfy
the Ginsparg--Wilson relation, and two gauge non-invariant
methods, one using different cutoffs for the fermions and the
gauge fields, and one using gauge fixing.  Open problems within
all three approaches are addressed.

\vspace{1pc}
\end{abstract}

\maketitle

\section{Introduction}

Let us start by recalling why the lattice formulation of 
vector-like gauge theories such as QCD is so successful.
Lattice QCD is manifestly gauge invariant, and this implies
that gauge degrees of freedom decouple in the regulated
theory, \ie\ at any momentum scale.  It is because of this
that it is possible to find lattice formulations of QCD for
which a hermitian, positive definite 
transfer matrix can be constructed, implying unitarity of
the regulated theory.  In addition, we have weak-coupling
perturbation theory (WCPT), which tells us how to take
the continuum limit, and what the scaling behavior of the
theory looks like.  In particular, because of WCPT and
asymptotic freedom, the number of quarks
in the theory is determined by the free quark action.
If, for instance, we use Wilson fermions, the fermion
spectrum is undoubled, with the doubler modes all having
energies of order $1/a$.

In contrast, the lattice formulation of chiral gauge
theories (ChGTs) is more complicated.  The fundamental reason
is the chiral anomaly, which implies that there is no
exact chiral ($\gamma_5$) invariance on the lattice
\cite{karsmi}.  The converse of this is the well-known
Nielsen--Ninomiya theorem, which says that a local lattice
theory with exact $\gamma_5$ invariance is in fact vector-like
\cite{nienim}: the theory has fermion doublers, and the
fermions that emerge in the continuum limit transform in a
vector-like representation of the (would-be) chiral symmetry group.
(Non-local theories, such as those using SLAC \cite{slac} 
or Rebbi \cite{rebbi} derivatives, have even more severe problems
\cite{karsmi2,bodkov,pel}.)

This implies that in order to ``keep a symmetry chiral,"
(and have each fermion produce its contribution to the chiral
anomaly), chiral symmetry has to be broken on the lattice.
If the chiral symmetry is to be gauged, as in ChGTs, this 
means breaking gauge invariance, and we have to worry about
the gauge degrees of freedom (\gdofs).

An example will illustrate the importance of this issue.
Throughout this talk, I will
assume that all fermions coupling to the gauge fields are left
handed (LH).  (All RH fermions can be written as LH through
charge conjugation in four dimensions.)  Remove the doublers
with a momentum-dependent Wilson mass term \cite{wil}
\begin{equation}
-\frac{r}{2}(\psibar^n_R(x)\Lapl\psi^c_L(x)+{\rm h.c.})\,,
\label{WILSON}
\end{equation}
where $\psi^n_R$ is a gauge-neutral spectator fermion introduced
solely in order to give the doublers of the 
charged LH fermion field $\psi^c_L$ a mass.
This term breaks gauge invariance, whether we make the laplacian
covariant or not, and hence, for simplicity, I will not do so.
After a gauge transformation $\psi^c_L\to\Phi\psi^c_L$, this becomes
\begin{equation}
-\frac{r}{2}(\psibar^n_R(x)\Lapl(\Phi(x)\psi^c_L(x))+{\rm h.c.})\,.
\label{WILSONGAUGED}
\end{equation}
At this point $\Phi$ is a random field, and, for $r$ not too
large, we expect that $\langle\Phi(x)\rangle=0$; hence
\begin{equation}
m_{\rm doubler}\sim r\langle\Phi(x)\rangle=0\,.
\label{DMASS}
\end{equation}
$\Phi$ represents the \gdofs, and we see that it couples to
the fermions because the Wilson term breaks gauge invariance.
(Indeed, the Wilson term leads to the correct expression for the triangle
diagram for smooth gauge fields \cite{karsmi}.)  The
dynamics of this coupling has drastic consequences for the
fermion spectrum: the doublers come back, and the gauge theory
is vector-like!  (See \eg\ ref. \cite{pet} for large $r$.)

If we supply dynamics for $\Phi$ to give it a vacuum 
expectation value $v=\langle\Phi\rangle\ne 0$, we get
\begin{equation}
m_{\rm doubler}\sim rv\sim m_{\rm gauge\ field}\,,
\label{MASSVEV}
\end{equation}
and the theory is still not really chiral, because $v$ sets
the scale of both the doubler and gauge-field masses.
(This is essentially the idea behind mirror fermions \cite{mon}.)
Clearly, we need more sophisticated dynamics for $\Phi$
if an approach like this is to work.  We have 
in principle two options:\footnote{Not everyone agrees that there
are only these two options.  See \eg\ ref.~\cite{cre} and refs.
therein.}\\
{A.} Use ordinary chiral symmetry.  If this is gauged,
gauge invariance has to be broken on the lattice, and the \gdofs\ 
couple to the fermions.  Therefore the dynamics of the \gdofs\
has to be controlled such that disasters as in our example
above do not happen.  Clearly, it is {\it not} enough to
give a prescription for the fermion determinant: we need the
``back reaction" {\it on} the fermions in order to find out whether 
or not the fermion spectrum remains chiral.
Realizations of this approach will be reviewed in sect. 3.\\
{B.} Change chiral symmetry on the lattice to make
it compatible with exact gauge invariance {\it and}
doubler masses.

That the latter option can be realized was actually discovered
rather recently, and came somewhat as a surprise.  Modify the
axial transformation rule of $\psi$, but not $\psibar$,
by a term of order $a$ \cite{luesym}:
\begin{equation}
\delta\psi=T\gamma_5(1-a{\cal O})\psi\equiv T\ghat\psi\,,\ \ \ 
\delta\psibar=\psibar\gamma_5T\,,\label{MODAXIAL}
\end{equation}
where $T$ are the generators of some Lie group, and ${\cal O}$ is
an operator of mass dimension one.  Choose ${\cal O}$
such that, in the free theory, $a{\cal O}\psi=0$ for $p=0$, but 
$a{\cal O}\psi=2$ for $p=(\pi/a,0,0,0)$ \etc.  This
changes the transformation of $\psi$ into
\begin{eqnarray}
\delta\psi(p=0)&=&T\gamma_5\psi(p=0)\,,\label{MOMAXIAL}\\
\delta\psi(p=\pi/a)&=&-T\gamma_5\psi(p=\pi/a)\,.\nonumber
\end{eqnarray}
This makes it possible to give modes at $p=\pi/a$ a
mass, without violating this symmetry.  This
new version of chiral symmetry has to satisfy two more
requirements.  First, we wish the symmetry group to be
the same as that of the continuum theory.  We therefore
need
\begin{equation}
\ghat^2\equiv[\gamma_5(1-a{\cal O})]^2=1\,,\label{OONE}
\end{equation}
which is the same as stating that ${\cal O}$ satisfies
the Ginsparg--Wilson (GW) relation \cite{ginwil}
\begin{equation}
\{{\cal O},\gamma_5\}=a{\cal O}\gamma_5{\cal O}\,.
\label{GINSWILS}
\end{equation}
Second, we need a fermion action invariant under this
symmetry. To this end, we choose ${\cal O}=D$, with $D$ a 
gauge-covariant
lattice Dirac operator satisfying eq.~(\ref{GINSWILS}),
and invariance under eq.~(\ref{MODAXIAL}) then follows.
In this talk, I will choose $D$ to be the overlap
operator of ref.~\cite{neu}.  The application of this
idea to ChGTs is the topic of the next section.
Note that the transformation rules of $\psi$ and $\psibar$
are not related.  This is no problem in the euclidean 
formulation, but raises the question how this would
carry over to a canonical formalism.

\section{Gauge invariant approach}
\subsection{Chiral fermion determinant}

With a Dirac operator obeying the GW relation, chiral 
projections of the fermion fields can be consistently 
defined as \cite{lueab,nied}
\begin{eqnarray}
\psi_L&=&\Phat_L\psi=\frac{1}{2}(1-\ghat)\psi\,,
\label{PROJECTORS} \\
\psibar_R&=&\psibar P_R=\psibar\frac{1}{2}(1+\gamma_5)\,,
\nonumber
\end{eqnarray}
and the action for a ChGT is then latticized as
$S=\sum\psibar_RD\psi_L$, which is  consistent because
$D\ghat=-\gamma_5D$.  The fermion
determinant is defined (up to a phase) by expanding $\psi_L(x)
=\sum_i c_i v_i(x)$ resp.
$\psibar_R(x)=\sum_j\vbar_j(x)\cbar_j$ on complete sets of 
states $\{v_i\}$ resp.  $\{\vbar_j\}$, with 
$\ghat v_i=-v_i$, $\vbar_j\gamma_5=\vbar_j$, and
$c_i$, $\cbar_j$ Grassmann variables:
\begin{eqnarray}
\det\,M&=&\int\prod_i dc_i\prod_jd\cbar_j\,e^{-\cbar_kM_{k\ell}
c_\ell}\,,\label{DET} \\
M_{k\ell}&=&\vbar_kDv_\ell\,. \nonumber
\end{eqnarray}
The overlap-Dirac operator is
\begin{eqnarray}
D&=&1+\gamma_5\epsilon(H)\,,\label{OVERLAPDIRAC}\\
H&=&\gamma_5(-1+D^{\rm massless}_{\rm Wilson})=H^\dagger\,,
\nonumber
\end{eqnarray}
with $D^{\rm massless}_{\rm Wilson}$ the standard Dirac--Wilson
operator with vanishing bare mass, and $\epsilon(H)$ is the sign
of $H$.  It follows that \cite{nar,kikyam}
\begin{equation}
\epsilon(H)=-\gamma_5(1-D)=-\ghat\Rightarrow M_{k\ell}=
2\vbar_kv_\ell\,.\label{EPS}
\end{equation}
Hence, $\{v_i\}$ ($\{\vbar_j\}$) is a complete set of states
of $H$ ($H'=\gamma_5$) with $H>0$ ($H'>0$), and we can make
contact with the overlap definition of the fermion determinant
\cite{narneu}
\begin{equation}
\det\,M\propto\langle 0-|0+\rangle_U\,,\label{OVERLAP}
\end{equation}
where $|0+\rangle_U$ ($|0-\rangle$) is the many-body ground state
of $-H$ ($-H'$). In fact, such a relation can be established
for any $D$ satisfying the GW relation \cite{nar}, but $D$ as
defined in eq.~(\ref{OVERLAPDIRAC}) is local if the gauge field
is restricted to satisfy 
\begin{equation}
|\tr(1-U_{\rm plaquette})|<\epsilon\,,\label{LOCALITY}
\end{equation}
with $\epsilon<1/30$ \cite{herjanlue}. 

This definition of the fermion determinant is not complete.
There is an arbitrariness in the choice of basis $\{v_i\}$,
which depends on the background gauge field $U$, because
$\ghat$ depends on $U$.  This leads to a phase ambiguity in the
determinant, which, in the overlap formulation, corresponds to
the phase ambiguity in the choice of $|0+\rangle_U$.  
The choice of phase has to break gauge invariance, so that
each fermion in a complex irreducible representation
will contribute its share of the chiral anomaly
(even if the complete fermion representation is anomaly free).
The question is whether this phase can be chosen
such that gauge invariance is preserved for anomaly-free
theories.

\subsection{Phase choice --- overlap}

In the original overlap formulation, the phase was determined
{\it \`a la} Wigner--Brillouin by choosing $\null_1\langle 0+|
0+\rangle_U$ real and positive (so that only a $U$-independent
ambiguity remains) \cite{narneu}.  This phase choice breaks
gauge invariance also for an anomaly-free fermion spectrum.
Therefore, the key question
is whether the resulting unphysical interactions between fermions
and \gdofs\ alter the fermion spectrum, as explained in 
section 1.

The issue of gauge invariance has been studied in detail in a
two-dimensional $U(1)$ model with four LH fermions of charge $1$
and one RH fermion of charge $2$ (which is anomaly free in $d=2$)
\cite{narneutwod,izunis}.  The most complete investigation in
this context is that of ref.~\cite{izunis}, and I will highlight
here some of the conclusions of this work.  Starting from some
smooth gauge field $U_\mu(x)$ in a finite volume $L\times L$,
consider the singular gauge field 
\begin{equation}
U^{(s)}_\mu(x)=\cases{e^{2\pi ic_\mu}U_\mu(x)\,,&
$x_\mu=x_\mu^0\,,$\cr
U_\mu(x)\,,&$x_\mu\ne x_\mu^0\,,$\cr}
\label{SING}
\end{equation}
with $c_\mu$ arbitrary constants, for some $x_\mu^0$.  
This gauge field is gauge
equivalent to the smooth field 
\begin{equation}
U^{(u)}_\mu(x)=e^{2\pi ic_\mu/L}U_\mu(x)\,.\label{SMOOTH}
\end{equation}
Therefore, in this anomaly-free theory, the fermion determinant
should be the same for both configurations if it were to be
gauge invariant, for all $c_\mu$.  However, ref.~\cite{izunis}
found that the imaginary part, ${\rm arg}\,\det\,M$,
changes by $O(1)$ for $c_\mu>1/4$, ${\it also}$ in the continuum
limit, $1/L\to 0$. 

An equilibrium ensemble of lattice gauge fields contains mostly
fields with many singularities such as the one described above.
While no systematic classification exists, I expect
that this lack of gauge invariance is a generic property of
lattice gauge fields, and not a ``set-of-measure-zero" problem.
(Note that ref.~\cite{narneutwod} only
considered gauge fields with singularities on the boundary.)
In fact, the class of gauge fields considered above includes fields
with vanishing field strength, but non-trivial Polyakov
loops.  On the $U(1)$ trivial orbit (with trivial Polyakov loops) 
it was proven that the determinant is gauge invariant in two dimensions 
\cite{narneu}, but this is a very special case, and only valid
in two dimensions.  It was shown that even on the
trivial orbit the overlap
definition with WB phase choice is equivalent to a theory with
domain-wall fermions in a waveguide, with both doublers 
and ghosts appearing
dynamically on the waveguide boundaries \cite{golshaov}.  While
the doubler- and ghost-contributions to the determinant may
cancel in two dimensions on the $U(1)$ trivial orbit, 
this is certainly not the case for general gauge fields in four 
dimensions.

In summary, the WB phase choice is not gauge invariant even
for anomaly-free theories, and there is good reason to expect that
this damages the chiral nature of the fermion spectrum.  Note
again that, in order to probe this issue, the back reaction of the
gauge field (in particular the \gdofs) has to be considered.

\subsection{L\"uscher's phase choice}

A different way of choosing the phase of the determinant was
proposed by L\"uscher \cite{lueab,luenonab} (see also
ref.~\cite{fujetal}).  Under an infinitesimal deformation of
the gauge field $\delta U_\mu(x)=i\eta_\mu(x)U_\mu(x)$ we have
\begin{equation}
\delta\log\det\,\!\!M\!=\!\Tr(D^{-1}\!P_R\delta D\Phat_L)
+\!\sum_{x\mu\,a}\eta^a_\mu j^a_\mu(x),\label{DEFORM}
\end{equation}
where the first term comes from the variation of the action.
The second term, equal to $\sum_{i,x}v^\dagger_i(x)
\delta v_i(x)$ and defining the current $j^a_\mu(x)$,
comes from the measure ($\eta_\mu=\eta_\mu^a T^a$ with
$T^a$ the gauge-group generators).  Specializing to a gauge
variation $\eta_\mu=\nabla_\mu(U)\omega$, with $\nabla_\mu(U)$
the gauge-covariant forward difference operator, this yields
\begin{eqnarray}
\delta\log\det\,M&\!\!=\!\!&i\sum_x\omega^a\left[
{\cal A}^a(x)-(\nabla^*_\mu j_\mu)^a(x)\right]\,,
\nonumber\\
{\cal A}^a&\!\!=\!\!&-i\,\tr(\ghat T^a)\,,\label{TRANSFORM}
\end{eqnarray}
with $\nabla^*$ the covariant backward difference operator.
The current $j^a_\mu$ is determined by the choice of measure%
\footnote{which, in turn, uniquely specifies the measure \cite{lueab}}.
What one would like to do, is to choose the measure such that
the determinant is gauge invariant, \ie\ such that
\begin{equation}
(\nabla^*_\mu j_\mu)^a(x)={\cal A}^a(x)\,.\label{ANOMCANC}
\end{equation}
This is, however, not the only requirement.  $j^a_\mu(x)$
has to be local and gauge covariant, so that the equations
of motion for the gauge field (derived with the help of
eq.~(\ref{DEFORM})) are local and gauge covariant.  
Furthermore, $j^a_\mu(x)$ has to satisfy an integrability
condition.  Consider a family of gauge fields $U_\mu(x;t)$
with $0\le t\le 1$ interpolating between two gauge fields
$U_\mu(x;0)$ and $U_\mu(x;1)$.  Evidently, we have
\begin{equation}
\log\frac{\det\,M(U(t=1))}{\det\,M(U(t=0))}
\!=\!\!\!\int_0^1\!\!\!\!\! dt\,\partial_t\log\det\,\!\!M(U(t)).
\label{INTCOND}
\end{equation}
The rhs of eq.~(\ref{INTCOND}) must therefore be
independent of the path $U_\mu(x;t)$, when expressed in terms of
the current $j_\mu^a$, using eq.~(\ref{DEFORM})
with $\eta_\mu=\partial_tU_\mu U^{-1}_\mu$.  This
defines the integrability condition on the current
$j_\mu^a$.

It is useful to consider the mathematical problem of the
existence and construction of a current satisfying all these
conditions first in the classical continuum limit.  For $a\to 0$,
\begin{eqnarray}
{\cal A}^a(x)&\propto& d^{abc}\epsilon_{\mu\nu\rho\sigma}
F_{\mu\nu}^b(x)F_{\rho\sigma}^c(x)\,,\label{CONTL}\\
d^{abc}&=&\tr(\{T^a,T^b\}T^c)\,,\nonumber
\end{eqnarray}
and eq.~(\ref{ANOMCANC}) reduces to the well-known 
(local) anomaly-cancellation condition of the continuum:
It is well-known that a current satisfying all conditions
cannot be constructed unless $d^{abc}=0$.  (If $d^{abc}=0$,
we can choose $j^a_\mu(x)=0$ in the classical continuum
limit.)  We recover the standard anomaly condition from
the lattice (as we would from any of the other proposals
discussed in this review!).  The key question is now whether
$d^{abc}=0$ is also {\it sufficient} on the lattice.

For a generic lattice definition of the fermion
determinant which agrees with the continuum for $a\to 0$,
the answer is {\it no}.  In general, one has, in a theory 
with $d^{abc}=0$ \cite{shalat},
\begin{equation}
\delta_\omega\log\det\,M=O(aA,ap)=O(1)\,,\label{VAR}
\end{equation}
because $aA$ and $ap$ (with $U={\rm exp}(iaA)$ and $p$
the momentum scale) are not small for typical lattice gauge 
fields.

In the case at hand, however, we know much more about the
structure of the anomaly on the lattice, ${\cal A}^a(x)$.
Using that $D$ satisfies the GW relation (and hence
$\ghat^2=1$), it can be proven that ${\cal A}^a(x)$ is
related to a topological field $q(z)$ in $4+2$ dimensions
($z=(x;s,t)$ with $s,t$ the extra coordinates\footnote{The 
two extra dimensions are continuous.}), \ie\ 
\begin{equation}
\int ds\,dt\sum_x\delta q(z)=0\,, \label{TOP}
\end{equation}
for any (local) deformation of the (six-dimensional)
gauge field \cite{luenonab}.  This is basically a  
lattice version of the relation between the non-abelian
anomaly in $d$ dimensions and the abelian anomaly in
$d+2$ dimensions through the descent equations
\cite{stozum}.  For the four-dimensional abelian theory
${\cal A}(x)$ is itself a topological field \cite{lueab}.

L\"uscher then proved the following theorem \cite{luenonab}:\\ \\
{\it $q(z)=$ total derivative $\Leftrightarrow$ There exists a current 
$j_\mu^a$ with all the desired properties}\\ \\
(modulo global obstructions, because the differential 
version of eq.~(\ref{INTCOND}) was used; see below).
The question thus reduces to whether
$d^{abc}=0$ is enough to ensure that $q(z)$ is a total
divergence.

In the abelian case the answer is, in fact, yes 
\cite{lueab,lueanom}, in theories in which there is an
even number of fermions for each odd charge.  
(For a reformulation of this choice
of phase in overlap language in the abelian case, see 
ref.~\cite{neuab}.)

In the non-abelian case there is, as of now, no non-perturbative
answer.  The answer is again yes to all orders in an expansion
in $a$ \cite{luenonab}, and, even stronger, to all orders in
an expansion in the gauge coupling $g$ \cite{luepert,suz}.

The approach of ref.~\cite{suz} is different, and deals with
the consistent anomaly $\hat{\cal A}$ instead of the covariant
anomaly ${\cal A}$ (eq.~(\ref{TRANSFORM})), without {\it a priori}
specifying action or measure.  $\hat{\cal A}$
is the variation of the effective action $\log\det\,M$,
and satisfies the Wess--Zumino consistency condition \cite{weszum}
\begin{equation}
\hat{\cal A}=\delta_{\rm BRST}\log\det\,M
\Rightarrow \delta_{\rm BRST}\hat{\cal A}=0\,.\label{WZ}
\end{equation}
It is then proven for the non-abelian case that, if $\hat{\cal A}$
is local and smooth and $d^{abc}=0$, $\hat{\cal A}$ can be
written as the BRST variation of a local and smooth functional
\begin{equation}
\hat{\cal A}=\delta_{\rm BRST}{\cal B}\,,\label{SOL}
\end{equation}
which can therefore be interpreted as a counter term.  The
proof is again given to all orders in $g$.

In the abelian case, the condition $d^{abc}=0$ reduces to
the requirement that $\sum_i e_i^3=0$, where the sum is over
all (LH) fermions (each with charge $e_i$) in the theory.
It was possible to complete the construction of  
the gauge-invariant determinant for the anomaly-free case
because it was possible to give a complete description of the
(admissible; see below) gauge-field configuration space
in finite volume (with periodic boundary conditions)
\cite{lueab}.  In fact, in finite volume one can also address
the issue of global obstructions, and, 
an additional condition for integrability 
is that there be an even number of
fermions for each odd charge.  A similar analysis
in the non-abelian case is much harder, and the program of 
non-perturbatively constructing a gauge-invariant fermion 
determinant for anomaly-free ChGTs has not been
completed.

In the abelian case, it was found that the gauge fields 
have to be admissible, that is, the gauge fields have
to satisfy the constraint
\begin{equation}
|1-U_{\rm plaquette}|<\frac{\pi}{3}\label{ADM}
\end{equation}
in order to avoid topological fields which are not
a total divergence, even in the anomaly-free case \cite{lueab}.
This constraint is independent from, but implied by,
the constraint eq.~(\ref{LOCALITY}), which guarantees
locality of the overlap-Dirac operator.  Obviously, this
constraint will have to be taken into account in 
numerical simulations of an abelian ChGT using this approach. 

In any numerical simulation of a ChGT, the most severe problem
will be the fact that a euclidean chiral determinant has a complex
phase.  This is a property inherent to ChGTs, and thus cannot
be avoided, irrespective of the details of the lattice
formulation.  A problem more specific to the gauge-invariant 
approach is the
need for a very accurate approximation to $\ghat
=-\epsilon(H)=-H/\sqrt{H^2}$.  Since the
properties of $\ghat$ play a crucial role in the proof of
gauge invariance, the situation is unlike that
in QCD with overlap fermions, where an approximate $\ghat$
(or, equivalently, an operator $D$ satisfying GW only approximately)
breaks chiral symmetry, but not gauge invariance.  Here,
an approximate $\ghat$ breaks gauge invariance, and
in such a situation the \gdofs\ couple to the physical sector.
One expects that for a very good approximation of $\ghat$
the phase of the determinant should be very close to its exact
value for any lattice gauge field.  On the other hand, in terms of a 
phase diagram, breaking gauge invariance implies moving away
from the gauge-invariant subspace (in particular from the
gauge-invariant critical point), and it would be interesting 
to understand how the decoupling of \gdofs\ takes place
with successively better approximations to $\ghat$.

One systematic way of approximating the overlap-Dirac operator
$D$, and hence $\ghat$, is through domain-wall fermions
in the limit of infinite fifth dimension \cite{kap,shadwf,vra}.
The connection between domain-wall fermions and the overlap-Dirac
operator in the context of L\"uscher's construction of ChGTs
has been considered in ref.~\cite{aoykik}.  It was proven that,
for smooth gauge fields (\ie\ in the limit $a\to 0$) and for
$d^{abc}=0$, the imaginary part of L\"uscher's chiral fermion
determinant is related to the $\eta$ invariant
\cite{eta}:
\begin{equation}
{\rm Im}\,\log\frac{\det\,M(U(t=1))}{\det\,M(U(t=0))}
=\pi\eta\,,\label{ETA}
\end{equation}
where formally $\eta=\sum_{\lambda\ne 0}\frac{\lambda}{|\lambda|}$,
with $\lambda$ the eigenvalues of the five-dimensional
Dirac operator
$D_5=i\gamma_5\partial_5+i\Dsl(U(x,t)).$
This is precisely the 
domain-wall Dirac operator with Shamir's boundary conditions
\cite{shadwf} (with a continuous fifth dimension).\footnote{In 
the continuum, the relation between
domain-wall fermions and the $\eta$ invariant was established
in ref.~\cite{kapsch}.}  However, at present this approach
has not yet led to a gauge-invariant definition of the determinant
through domain-wall fermions.

\subsection{Applications}

While the program of defining a gauge-invariant chiral fermion
determinant on the lattice starting from the symmetry 
eq.~(\ref{MODAXIAL}) is not complete, and numerical simulations are
certainly difficult, this approach has led to some 
interesting theoretical developments.

First, as I have mentioned already above, the program was completed
in perturbation theory \cite{luepert}.  This means that, to all 
orders in perturbation theory, an anomaly-free ChGT
can be put on the lattice without violating gauge invariance.
In perturbation theory, the locality constraint eq.~(\ref{LOCALITY})
is automatically satisfied, and therefore the theory is also local.

This lattice formulation of anomaly-free ChGTs thus provides
a manifestly gauge-invariant four-dimensional perturbative
regulator for ChGTs.  As such, it should
be of practical interest for instance in the calculation of
perturbative corrections in the electro-weak sector of the
Standard Model and for electro-weak hadronic matrix elements.
A property which is not 
manifest in this lattice formulation is unitarity.  With the program
being complete in perturbation theory,
it should be possible to establish unitarity of anomaly-free
gauge theories at least to all orders perturbatively.

Another interesting application of the construction of 
abelian ChGTs is that it can be extended to
include the full electro-weak ($U(1)_Y\times SU(2)_L$) sector
of the Standard Model \cite{kiknak}.  The essential ingredient is
that the gauge-fermion sector of the Standard Model
is chiral only because of the presence of $U(1)_Y$ (all
representations of $SU(2)$ are pseudo-real).  This makes it
possible to carry over the non-perturbative construction of
abelian ChGTs of ref.~\cite{lueab}, as was done in 
ref.~\cite{kiknak}.\footnote{Since
$SU(3)$ does not share this property, the construction does not
extend to $SU(3)_{\rm color}$.} The construction was carried out
in infinite volume, \ie\ without consideration of global 
obstructions to the integrability condition (\ref{INTCOND}). 

There may exist global obstructions to the integrability
condition eq.~(\ref{INTCOND}) in its integral form.  Even if
this condition is satisfied in its local (differential) form,
it may still be that the rhs of eq.~(\ref{INTCOND}), when
expressed in terms of eq.~(\ref{DEFORM}), is not
independent of the path in general, 
or equivalently, that a closed path
exists for which the rhs does not vanish.  In fact, such a
closed path exists
for a single chiral fermion in the fundamental
representation of $SU(2)$ \cite{barcam}.
Hence, such a theory has a global anomaly.  Moreover,
ref.~\cite{barcam} showed that this obstruction
of the integrability condition is precisely the way in which
the Witten anomaly \cite{wit} manifests itself in this
regularization of an $SU(2)$ ChGT, and also
considered other $SU(2)$ representations.

While this is a very nice result, it is not the same as a
complete classification of all global obstructions in (this
definition of) ChGTs.  A complete classification would
be necessary in order to construct lattice ChGTs in a
finite volume, and would entail a study of {\it all}
possible closed paths in the relevant gauge-field space.
Such an endeavor is obviously much more ambitious.
It should be said in this respect that the other proposals 
discussed
in this talk do not lend themselves to a systematic analytic
study of global obstructions.

To close this section, I mention some work on the family-index
theorem for the overlap operator \cite{ada}.  While 
potentially interesting, thus far only the necessary
requirement that $d^{abc}=0$ in a lattice definition of
a ChGT has been recovered.

\section{Gauge non-invariant approach}

I now return to ``option A" in the Introduction, the
construction of lattice ChGTs using ordinary chiral symmetry,
which leads us to consider gauge non-invariant approaches.  

As we have seen in the Introduction, great care has to be
taken with respect to the dynamics of the \gdofs\ in any
approach which is not gauge invariant on the lattice.
We start again by considering Wilson fermions.  The Wilson
mass term is not gauge invariant, as shown in 
eq.~(\ref{WILSONGAUGED}), where the \gdofs\ $\Phi$ have been 
made explicit.  The lattice action {\it is} invariant under
the extended symmetry
\begin{eqnarray}
\psi^c_L(x)&\to& h_L(x)\psi^c_L(x)\,,\label{EXTGAUGE}\\
\psi^n_R(x)&\to& h_R\psi^n_R(x)\,,\nonumber\\
\Phi(x)&\to& h_R\Phi(x)h^\dagger_L(x)\,.\nonumber
\end{eqnarray}
The local symmetry $h_L(x)$ is, of course, 
not the symmetry of the ChGT we are
after, because it transforms both physical degrees of freedom
($\psi^c_L$) as well as unphysical ones ($\Phi$).
Our task is to control the dynamics of $\Phi$ such that

1) $\Phi$ {\it decouples from the physical spectrum;} 

2) $\langle\Phi\rangle=0$, \ie\ {\it the gauge group is unbroken;}

3) {\it there are undoubled (massless) charged LH fermions,
and no charged RH fermions.} \\
The latter two requirements
are well-defined in the theory in which the gauge field is constrained
to be longitudinal,  keeping the interactions
between the fermions and the \gdofs\ (\cf\ eq.~(\ref{WILSONGAUGED})).
The transverse degrees gauge fields are controlled by the gauge
coupling $g$, and, as in QCD, are not expected to change the chirality
of the fermions to which they couple.

Before I review two specific proposals on how to do this, let us
make one more general observation.  One may define a {\it charged}
RH field
\begin{equation}
\psi^c_R=\Phi^\dagger\psi^n_R\,,\ {\rm so\ that}\ \psi^c\to h_L\psi^c
\,,\label{CHRH}
\end{equation}
and one can then
find out what states are created by $\psi^c_L$ and $\psi^c_R$ 
from the charged propagator
\begin{equation}
S^c(p)=\langle\psi^c(p)\psibar^c(x)\rangle e^{ipx}\,.\label{CHPROP}
\end{equation}
It was shown in ref.~\cite{shanogo}
that if the inverse of this propagator has a continuous
first derivative, the Nielsen--Ninomiya theorem applies, and hence,
that species doubling occurs, rendering the gauge theory non-chiral.
Therefore, for any given proposal, we need to find
singularities in the charged fermion propagator, and understand their
nature.  This will be a key point in my discussion of the two concrete
proposals below.

\subsection{Two cutoff method}

One proposal is to regularize a ChGT with different cutoffs for the
fermions and the gauge fields \cite{gocetal,hoo,hersun,bod,hsu,sla,boretal}.
This can be done in many different ways.  Here I will discuss the
implementation in which the gauge fields live on a lattice with
spacing $b$, and the fermions on a lattice with spacing $f\ll b$
\cite{hersun,bod}.
The gauge fields $U$ on the coarse lattice are interpolated smoothly
and covariantly
to a gauge field on the fine lattice on which the fermions live:
\begin{equation}
u_\mu(y)=u_\mu(y;U_\nu(x))\,,\ \ \ y\in H(x)\,,
\label{INTERPOL}
\end{equation}
with $H(x)$ the coarse-lattice hypercube located at $x$, and
$y$ resp. $x$ labeling the sites on the fine resp. coarse lattice.
The fermions are then coupled to this interpolated field $u$.
The Wilson term is still not gauge invariant, but roughly the idea
is that the interpolated field now is always smooth on the scale
$f$, so that the \gdofs\ are innocuous.  

One integrates out the fermions, removing divergences (which
may be gauge non-invariant!) with counter terms.  It was then
proven \cite{hersun,bod} in perturbation theory that the
remaining violations of gauge invariance are of order $f^2/b^2$,
if the theory is anomaly free.
(A similar statement applies to fermionic correlation functions.)
The continuum limit is then taken by taking the limit
$f/b\to 0$ first, and then $b\to 0$.

In fact, it has been suggested that one may not need to take
the limit $f/b$ first, but that it might suffice to take the ratio
$f/b$ small enough.  This suggestion calls on an argument by
Foerster, Nielsen and Ninomiya \cite{fnn}, who proved that
a small explicit bare mass term added to the (lattice)
Yang--Mills lagrangian is irrelevant with respect to the
standard Yang--Mills critical point at $g=0$, and therefore the
``small" breaking of gauge invariance by this mass term
does not change the
long-distance behavior of the theory.  (For a more complete
discussion of the relevant phase diagram, see ref.~\cite{frashe}.)
However, a very important difference between the case studied
in ref.~\cite{fnn} and the case at hand is that the $O(f^2/b^2)$
(and higher-order) terms in the fermion determinant are non-local.
Because of this I believe that it is unlikely that
a similar argument applies to the case at hand.  

There is another caveat concerning the issue of locality:
a smooth, gauge-covariant interpolation cannot be local \cite{shalat}!
It is sufficient to demonstrate this in the context of a 
two-dimensional $U(1)$ theory.  Consider a plaquette with
links $U_1(x)=e^{i(\pi-\epsilon)}$, $U_2(x+\hat 1)=
e^{i(\pi-\epsilon)}$, $U_1(x+\hat 2)=e^{i(-\pi+\epsilon)}$,
$U_2(x)=e^{i(-\pi+\epsilon)}$, where $\epsilon$ is small.
($\hat 1$ is the unit vector in the 1-direction, \etc.)
The plaquette equals $e^{-4i\epsilon}$, and the field strength
is $F_{12}=-4\epsilon$.  In order to interpolate, we define
a potential $A_\mu$ on the links of the coarse lattice by
\begin{equation}
U_\mu(x)=e^{iA_\mu(x)}\,,\ \ \ |A_\mu(x)|\le\pi\,.\label{GPOT}
\end{equation}
The smooth interpolation of refs.~\cite{gocetal,hersun,bod} for
this simple case is
\begin{eqnarray}
a_1(y)&=&A_1(x)(1-y_2)+A_1(x+\hat 2)y_2\,,\label{INTAB}\\
a_2(y)&=&A_2(x)(1-y_1)+A_2(x+\hat 1)y_1\,,\nonumber
\end{eqnarray}
leading to a field strength $f_{12}=\partial_1a_2
-\partial_2a_1=4\pi-4\epsilon\ne F_{12}$!  The reason for this
is that the winding number of our field around the plaquette does
not vanish ($\Pi_1(U(1))\ne 0$).  
This simple interpolation is not covariant.

This problem can be solved by first performing a gauge transformation
on $U_\mu$, so that after the transformation $U_1(x)=e^{-i\epsilon}$,
$U_2(x+\hat 1)=e^{-i\epsilon}$, $U_1(x+\hat 2)=e^{i\epsilon}$,
$U_2(x)=e^{i\epsilon}$.  Interpolating again, one finds a gauge
field $a'_\mu$ with $f'_{12}=-4\epsilon$, which does equal $F_{12}$.
(Note that $a'_\mu$ is not a gauge transform of $a_\mu$!)  Equivalently,
one may choose the field $A$ such that the winding number around the
plaquette vanishes (by choosing a different branch of the logarithm
in eq.~(\ref{GPOT})), in our example \eg\ $A_2(x)=3\pi-\epsilon$,
which would lead to $f_{12}=-4\epsilon$.  This can be done 
systematically \cite{hersunnl}, but it has to be done plaquette
by plaquette for each coarse-lattice configuration, and hence 
makes the interpolation procedure non-local.  For a different
suggestion, see ref.~\cite{bodnl}.  In four-dimensional
non-abelian gauge theories a similar problem occurs, because
$\Pi_3(G)\ne 0$ for any non-abelian group $G$.  

Numerical exploration of this idea is obviously demanding, because
large lattices are required for the fermions, even if the coarse
lattice is rather small.  Probably because of this reason, there
exists only one non-perturbative study to date addressing the
three questions below eq.~(\ref{EXTGAUGE}).  In ref.~\cite{herbou}
the interpolation approach was tested for a $U(1)$ model in
two dimensions, in the limit of vanishing gauge coupling.\footnote{%
For numerical studies restricted to the fermion determinant,
see ref.~\cite{sch} and refs. therein.}  In that
limit, the only interaction is that of the \gdofs\ with the
fermions.  Like the gauge field $U$, the field $\Phi(x)$ is
interpolated smoothly to a field $\phi=\phi(y;\Phi(x))$ with
$y\in H(x)$ on the fermion lattice,
and this interaction has the form
\begin{equation}
-\frac{r}{2}\sum_{y}
\left(\psi^n_R(y)\Lapl(\phi(y;\Phi(x))\psi^c_L(y)\!+\!{\rm h.c.}\right)\,.
\label{WYINT}
\end{equation}
This Wilson--Yukawa coupling is of the form
\begin{equation}
rq^2\delta(p-q+k)\,,\ \ \ k\sim\frac{1}{b}\ll\frac{1}{f}\,,
\label{WYCOUPL}
\end{equation}
with $p,q$ the fermion momenta, and $k$ the scalar momentum.
Because the scalar field lives on the coarse lattice, and is
smoothly interpolated to the fine lattice, it contains only
momenta less than or of order $1/b$, and the $\delta$ function
sets $p\approx q$.  This leads one to expect that for $p,q$ small
this coupling is irrelevant, so that this interaction will not
affect the physical modes, while it will be strong for large
$p,q$, so that it could potentially provide the doublers with
a mass.  This is not unlike a scenario envisaged long ago
\cite{eicpre}, which back then did not work \cite{golpetriv}
because all fields lived on the same lattice, leading to
strong couplings between the physical and doubler modes
through Wilson--Yukawa-like interactions.

The scalar and fermion spectrum of this model was studied
numerically on lattices of size up to $L/b=10$, with lattice-spacing
ratios up to $b/f=8$ in the quenched approximation, with no
further $\Phi$-dependent terms in the action (no hopping term), 
\ie\ $\Phi$ was
taken to be a random $U(1)$ phase.  Note that also in this case
the smooth, covariant interpolation is non-local, in the sense
described above.  To date, there have been no analytic studies
of this model.

First, one expects $\langle\phi\rangle$ to vanish ($\Phi$ is a
random field), and this is indeed what was found, within 
errors.  Then, it was found that the $\phi$ two-point function
could be fitted satisfactorily to a free scalar lattice
propagator for momenta $\le 1/b$, with a scalar mass 
$m_\phi\approx 1.7/b$.  This means that the unphysical 
degrees of freedom represented by $\phi$ are removed from the
long-distance physics of the model by acquiring a mass of order
the cutoff.  Note that the correlations leading to this mass
are entirely due to the non-local nature of the interpolation,
because the field $\phi$ would otherwise be a random field
on the scale of the coarse lattice!

Fermion propagators were investigated in great detail, and
here I will only be able to discuss part of the results.
As we have seen, one important probe is the charged fermion
propagator (\ref{CHPROP}), for which numerical results are shown
in fig. 1 (a very similar plot was obtained for the neutral
fermion propagator).
The data shown in the figure are obtained by first 
projecting $S^c_{RL}=P_RS^cP_L$ \etc., and then plotting the
inverse of that, as a function of $q=(q_t,0)$.  The apparent
poles at $q_t=\pi$ show up as a consequence of this way of
representing the data.  In fact, at least qualitatively,
the data is well described by the {\it ansatz}
\begin{equation}
(S^c)^{-1}=\sum_\mu i\gamma_\mu\sin{q_\mu}
(P_Rh(q)+P_L)+M(q)\,,\label{ANSATZ}
\end{equation}
with $M(q)$ a Wilson-like mass term, which renders the
doublers massive,\footnote{The doublers can have rather
non-trivial interactions in two dimensions, see ref.~\cite{herbou}.}
and $h(q)$ a function to be discussed below.
\begin{figure}[htb]
\vspace{-20pt}
\includegraphics[width=2.7truein,height=2.6truein]{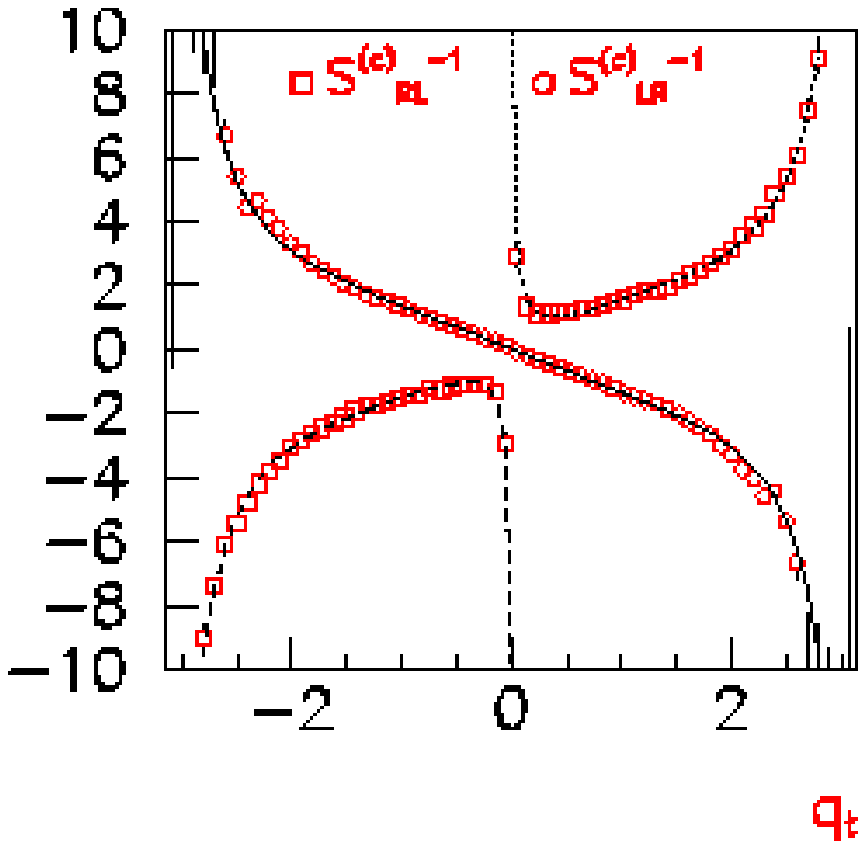}
\vspace{-35pt}
\caption{}{Components of the charged fermion propagator,
$(S^c_{RL}(q))^{-1}$ and $(S^c_{LR}(q))^{-1}$.
From ref.~\cite{herbou}.}
\vspace{-10pt}
\end{figure}
This {\it ansatz} contains a free LH charged fermion, which is
what we are aiming for.  (In fact, a similar {\it ansatz} describes
the propagator for the neutral fermion $\psi^n=\Phi\psi^c$
(not shown, \cf\ ref.~\cite{herbou}), for which shift symmetry
\cite{golpet} implies a form as in eq.~(\ref{ANSATZ}) with
$P_R$ and $P_L$ interchanged and with
$M(q)$ exactly the Wilson mass term, with only $h(q)$ not 
determined by this symmetry.)  Comparing this {\it ansatz}
with the figure, we find that
\begin{equation}
h(q)\sim\cases{1/q^2\,,&$q$ small\,,\cr 1\,,&
$q\sim(\pi,0),(0,\pi),(\pi,\pi)$\cr}\label{H}
\end{equation}
(an example of such a function is 
$h(q)^{-1}=\sum_\mu\sin^2{p_\mu}+(1-\prod_\mu\cos^2{(p_\mu/2)})^2$).
This pole is visible in the figure in the data for $(S^c_{RL}(q))^{-1}$.
This implies that the RH part of the inverse
charged propagator is singular near
$q=0$, the necessary condition for avoiding the Nielsen--Ninomiya theorem.
However, this singularity, if our {\it ansatz} is correct, 
represents a {\it pole} in the inverse propagator.  Through the
Ward identities for the symmetry (\ref{EXTGAUGE}) (with $h_L$
constant), this implies the
existence of a pole in the vertex function, because the charged RH channel
couples to the gauge field.  Such poles may in general lead to
unphysical contributions to the $\beta$-function of the full
theory \cite{bodkov,pel}.

I believe that the puzzle about the nature of this singularity
which apparently is present in the data should be resolved
in order to assess the viability of this approach to putting
ChGTs on the lattice.  While it does look that the model 
reviewed here contains an undoubled LH charged fermion,
it is not clear that there are no other light or massless
unphysical modes.  
It should be noted that it is quite possible that the behavior 
sketched here
is specific to two dimensions.  An analytic study, or a more
detailed numerical study, if feasible,
would obviously be helpful in interpreting the data appropriately. 
I will return to this approach toward the end of this section.

\subsection{Gauge fixing method}

An alternative gauge non-invariant approach keeps the fermions
and the gauge fields on the same lattice, but employs gauge
fixing in order to control the \gdofs\ \cite{bmrst,shagf,golshagf,vin}.
The central observation is that gauge fixing gives us the
possibility to control both the transverse {\it and} 
longitudinal gauge fields in lattice perturbation theory,
provided a suitable lattice action is chosen \cite{shagf}
(see below).  
In particular, gauge fixing suppresses fluctuations of the
longitudinal components (the \gdofs), and we will see that this 
makes it possible to keep the fermions chiral, 
unlike \eg\ the example discussed in the Introduction.
However, as we will see, in order to turn lattice perturbation
theory into a valid, systematic expansion scheme, non-perturbative
considerations are important. 

Again, I will consider Wilson fermions, with a momentum-dependent
mass term as in
eq.~(\ref{WILSON}) in order to remove doublers.  In the gauge-fixing
approach one adds to the naively discretized theory with charged
LH fermions and neutral RH fermions the terms
\begin{eqnarray}
&&-\frac{r}{2}(\psibar^n_R\Lapl\psi^c_L+{\rm h.c.})
+\tk(\partial_\mu A_\mu)^2 \label{GFAPPR} \\
&&+\kappa A_\mu^2+{\rm other\ counter\ terms}+{\rm ghosts}\,,
\nonumber \\
&&\tk=\frac{1}{2\xi g^2}\,, \nonumber
\end{eqnarray}
where $\xi$ is the gauge parameter, and $\kappa A_\mu^2$ 
is a mass counter term.  Note that the parameter $\tk$ controls
the longitudinal fluctuations just like $\beta=1/g^2$ controls the
transverse fluctuations.
Because the Wilson mass term breaks gauge invariance, there is no
BRST invariance on the lattice, and counter terms will be needed
in order to recover the Slavnov--Taylor identities in the continuum
limit \cite{bmrst}.  I will limit myself to the
abelian case, for which no ghost fields are needed \cite{bgsnogh}. 
Addition of the covariant gauge-fixing term in eq.~(\ref{GFAPPR})
makes the theory perturbatively
renormalizable, and this determines the form of the
counter terms: one simply has to add all possible terms up to 
mass dimension 4 consistent with the exact symmetries of the 
regulated theory.  The most important of these is the gauge-field
mass term; all others are of dimension 4 (because of shift
symmetries \cite{golpet,bmrst}).  To this point, we are just describing
what happens if one uses a gauge non-invariant perturbative
regulator.

Of course, for this to work non-perturbatively, the lattice
version of the gauge-fixing term proportional to $(\partial_\mu
A_\mu)^2$ has to be chosen such that the (non-perturbative)
phase diagram of the lattice theory has a critical point which
is described by this perturbative expansion.  One has to choose
the lattice gauge-fixing action such that the perturbative
vacuum $A=0$ (or, on the lattice, $U=1$) gives the dominant
contribution in a saddle-point approximation to the lattice
theory.  That this can be done was shown in ref.~\cite{golshagf}.
The idea is to choose a gauge-fixing term on the lattice
roughly of the form $\tk[(\partial\cdot A)^2+A^6]$, 
with the irrelevant $A^6$
term stabilizing the perturbative vacuum $A=0$, without
otherwise affecting the long-distance behavior.  (This shape
can be maintained to all orders in perturbation theory by
adjusting the counter terms.)  
The phase diagram which emerges is shown in fig. 2, where
only the $\kappa$-$\tk$ plane is shown \cite{shagf,golshagf,bgsred,bgls}
(see these refs. for other counter terms).  

Consider first the theory without fermions.
In a bounded region around the origin, \ie\ for small $\kappa$
and $\tk$, the gauge-fixing and mass terms do not change the
critical behavior of the theory: this region of the phase
diagram is filled by a Coulomb phase \cite{fnn}.  Since $\tk$ is small,
WCPT does not apply in this region.  For large $\tk$, WCPT
is applicable, and one finds a critical line where the gauge
field is massless by tuning the counter term $\kappa$
to a critical value $\kappa_c(\tk)$.  For
$\kappa$ too large, $m^2_A\propto\kappa-\kappa_c$ is positive,
and the gauge field acquires a mass.  For $\kappa$ too small,
$m^2_A<0$, which signals symmetry breaking.  Since in this
case $A_\mu$ acquires an expectation value $\langle A_\mu\rangle
\ne 0$, the broken symmetry is hypercubic invariance!  
\begin{figure}[htb]
\vspace{0pt}
\includegraphics[width=2.7truein,height=2.6truein]{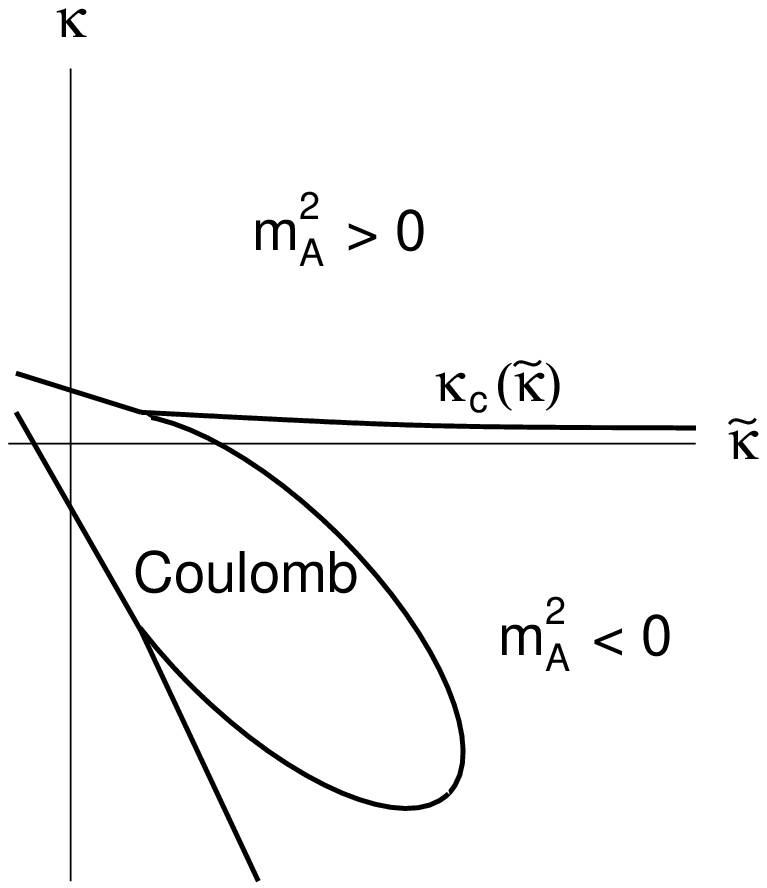}
\vspace{-30pt}
\caption{}{Sketch of the relevant part of the
phase diagram for the gauge-fixing approach.  For full details
on the complete diagram, see ref.~\cite{bgls}.}
\vspace{-20pt}
\end{figure}
Moreover, with our choice of irrelevant terms ($A^6$), this
transition is continuous.  To lowest order in WCPT,
the classical potential for the gauge field is (roughly)
\begin{equation}
V_{\rm cl}=\kappa A^2+\tk A^6\,.\label{VCLASS}
\end{equation}
Without the $A^6$ term, we would find a first order phase transition
going from $\kappa>\kappa_c$ to $\kappa<\kappa_c$ (where at tree level
$\kappa_c=0$), but with the $A^6$ term, the transition is
continuous.  We see
that the usual adjustment of the gauge-field mass counter term
corresponds, as desired, to a continuous phase
transition between two distinct phases above and below the
critical line $\kappa=\kappa_c(\tk)$.  For details,
see refs.~\cite{bgls,bgsred} (for an earlier informal
account, see ref.~\cite{bgsedin}).  A naive discretization
of $(\partial\cdot A)^2$ without the $A^6$-like term leads
to a very different phase diagram, without the desired
critical behavior \cite{shagf,bgls}.

We thus recover Maxwell theory (without fermions, thus far)
in two different ways: inside the Coulomb phase, and for large
$\tk$ near the critical line $\kappa=\kappa_c(\tk)$.  
The difference is that lattice perturbation theory only applies 
in the latter case, and this will be essential to our goal
of obtaining undoubled chiral fermions.  Inside the Coulomb
phase, the situation would be similar to the example in the
Introduction; the gauge-fixing term would not be strong enough
to control the longitudinal gauge field, \ie\ the \gdofs.

We now turn to the fermions.  First, since they are coupled to
the gauge field through the gauge coupling $g$, which is small,
one expects no qualitative change in the nature of the phase
diagram.  The issue is whether the ``back reaction" of the
gauge fields, in particular the \gdofs, is small as well.
Since this is now also governed by WCPT, one expects
that the fermion spectrum does not differ from that of the
classical continuum limit.  This is precisely what we find.
Because the danger resides in the dynamics of the \gdofs,
it is enough to consider (again, as we did in the two-cutoff
approach) only the interactions between the fermions and the
\gdofs.  As in eq.~(\ref{WILSONGAUGED}), we rotate
$\psi^c_L\to e^{i\lambda}\psi^c_L$, and we set $A_\mu=
\partial_\mu\lambda$, under which eq.~(\ref{GFAPPR}) goes
over into
\begin{eqnarray}
&&-\frac{r}{2}(\psibar^n_R\Lapl(e^{i\lambda}\psi^c_L)
+{\rm h.c.}) \label{GFGAUGED} \\
&&+\tk(\Lapl\lambda)^2+
\kappa(\partial_\mu\lambda)^2+\dots\,,\nonumber
\end{eqnarray}
where $\dots$ stand for the various $\lambda$ self interactions
coming from the lattice versions of the gauge-fixing, mass,
and other counter terms.  This model can be studied in WCPT
in $1/\sqrt{\tk}$
by rescaling $\lambda=\theta/\sqrt{2\tk}$ \cite{shagf,bgspt}.  The
following heuristic arguments describe what happens.

First, near $\kappa=\kappa_c$ the $\theta$ propagator goes 
like $1/p^4$, \ie\ the $\theta$ field has mass dimension zero.
Moreover, one finds that \cite{bgspt}
\begin{eqnarray}
\langle\Phi\rangle&=&\langle e^{i\lambda}\rangle\sim
(\kappa-\kappa_c)^\eta\,,\label{VEV} \\
\eta&=&\frac{1}{32\pi^2\tk}\,.\nonumber
\end{eqnarray}
This is analogous to the behavior of a near massless
scalar in two dimensions \cite{col}.  This means that at
$\kappa=\kappa_c$ the LH and RH symmetries ($h_L$ and
$h_R$ in eq.~(\ref{EXTGAUGE}))
are unbroken.  This is an important result: we are after
a ChGT with unbroken gauge symmetry, wherein the \gdofs\ $\Phi$
are to decouple completely!

Second, we may now expand the Wilson term in eq.~(\ref{GFGAUGED}),
obtaining
\begin{equation}
-\frac{r}{2}\psibar^n_R\Lapl\psi^c_L-
i\frac{r}{2}\frac{1}{\sqrt{2\tk}}\psibar^n_R\Lapl(\theta\psi^c_L)
+O(\frac{1}{\tk})+{\rm h.c.}\label{GFWILSON}
\end{equation}
The first term is the standard Wilson term, removing the fermion
doublers.  In addition, we see that the interaction terms between 
the fermions and the
field $\theta$ are all of dimension 5, and hence irrelevant!
They may lead to contact terms in fermionic correlation
functions, but they do not change the long-distance behavior
of the fermion fields.

Of course, a more rigorous argument is needed 
to confirm this picture in the presence
of the unusual infrared singular behavior of the $\theta$ field.
A study in one-loop resummed WCPT was performed in ref.~\cite{bgspt},
and detailed (quenched) numerical computations were done in 
ref.~\cite{bgsf,bgsred}.\footnote{Fermion loops appear at higher
order in WCPT, and are not expected to change our conclusions.}  
The numerical and analytic calculations
are in good quantitative agreement with each other, and confirm
the analysis described above.  We conclude that this theory 
contains undoubled chiral fermions, with the LH charged fermion
coupling to the gauge field, and the RH neutral fermion decoupling
in the continuum limit.  The \gdofs\ ($\theta$) decouple in the
continuum limit as well, as they should.  The gauge-fixing action 
gives the \gdofs\ the sophisticated dynamics alluded to in the
Introduction.  Since the transverse gauge field couples weakly
to the fermions, one does not expect any drastic change from
our conclusions in the full theory containing {\it both}
dynamical fermions and the full gauge field.

The central idea in this approach is to secure renormalizability even 
in the absence of gauge invariance.  The key ingredient is
gauge fixing, and one would therefore
expect universality with respect to the choice of lattice fermions.
This was indeed confirmed by a combined perturbative and numerical
study using domain-wall fermions instead of Wilson fermions
\cite{basde}.

The main issue in extending this method to non-abelian ChGTs is
the existence of Gribov copies.  Little work has been done in
this direction at present, and therefore I will not devote much
space to it here.  A promising approach appears to be to use the
non-perturbative gauge-fixing method proposed by Jona-Lasinio
and Parrinello and by Zwanziger \cite{jlpz}.  For speculative
thoughts on how to apply this method to ChGTs, see ref.~\cite{bglsdub}. 

\subsection{Comparison between gauge-fixing and two-cutoff methods}

As in subsection 3.1, we may again ask what happens to the charged
fermion propagator in the gauge-fixing approach.  The charged
RH fermion field is, as before,
\begin{equation}
\psi^c_R=\Phi^\dagger\psi^n_R=
e^{-i\theta/\sqrt{2\tk}}\psi^n_R\,,\label{RHCH}
\end{equation}
but now with $\theta$ describing the modes of the scalar field
$\Phi$.  Since this scalar field decouples from the fermions
for $a\to 0$, this composite operator describes exactly what
it shows: a multiparticle state composed of RH neutral fermions
and (higher-derivative) scalars in the continuum limit.  
The LH charged fermion
itself is a free field in the model of eq.~(\ref{GFGAUGED}),
and one finds for the LH, resp. RH components of the
charged fermion propagator \cite{bgspt,bgsedin}
\begin{equation}
P_LS^c(p)P_R\sim P_L\frac{1}{\psl}\,,\ 
P_RS^c(p)P_L\sim P_R\frac{1}{\psl}\,p^{4\eta}\,,\label{CHPROPGF}
\end{equation}
with $\eta$ given in eq.~(\ref{VEV}).
We see that, also in the gauge-fixing approach, the inverse 
charged propagator does not have a continuous derivative
at $p=0$. But now this singularity takes the form of a cut
occurring in the RH channel, as one would expect if $\psi^c_R$
creates a multiparticle state.  Because all these excitations
($\psi^n_R$ and $\theta$) are massless, the cut has a
branch point at $p=0$.

It is interesting to compare this with our interpretation
of what happens in the two-cutoff approach.  In both
approaches the unphysical scalars decouple, but in different
ways.  In the two-cutoff study, the \gdofs\ get a mass of
order the cutoff $1/b$, while in the gauge-fixing case,
they are massless at the relevant critical point, but have
no interactions with the physical sector.  One would 
expect that  
these two different ways of decoupling the \gdofs\ lead
to different singularities in the charged fermion propagator.
It is possible that the pole-like singularity exhibited
in fig. 1 is a finite-volume artifact, and would turn into
a branch point in infinite volume.  However, the existence
of a branch point at $p=0$ is difficult to reconcile with
scalar fields with a mass of order the cutoff.
Clearly, a better understanding is needed for the two-cutoff
method.

\section{Conclusion}

In this talk, I have covered what I consider the most
important developments since the last general review of this
field \cite{shalat}.  There has been a tremendous amount of
progress, and we appear to be much closer to satisfactory
constructions of lattice ChGTs.  However, the job is not done
yet.

Around the time of Shamir's 1995 review, 
the fundamental problem with regard to maintaining
the chiral fermion spectrum in the presence of interactions
with \gdofs\ had been well understood, and both the gauge-fixing
and the two-cutoff or interpolation methods have been proposed
in order to specifically address this issue.

More recently, L\"uscher has realized that it may even be possible
to formulate ChGTs on the lattice in a manifestly
gauge-invariant way, provided that (at least) the fermion
representation is anomaly free.  In this approach, the
Nielsen--Ninomiya theorem is circumvented by modifying
chiral symmetry on the lattice such that it yields the
usual chiral Ward identities in the continuum limit,
and yet does not lead to species doubling.   In order to
define ChGTs, the phase of the chiral determinant has to
be chosen so as to not violate gauge invariance.  Obviously,
the advantage of a manifestly gauge-invariant construction
is that the \gdofs\ decouple already from the physical
excitations at finite lattice spacing.

In order to show that ChGTs can be put on the lattice in
a gauge-invariant way, a complete classification of all
anomalies, both local and global, is necessary.  Such
a classification was accomplished for the case of abelian
ChGTs.  A crucial element is the admissibility constraint
eq.~(\ref{ADM}).  It is interesting to observe that, while this
constraint is not expected to change the universality class
in the weak-coupling limit \cite{lueab}, it does change the
strong-coupling behavior of this lattice version of ChGTs
relative to that of vector-like theories.  
For the non-abelian case, there is a conjecture
that the standard anomaly-cancellation condition $d^{abc}=0$
is sufficient for removing all local anomalies.  Little
is known at present about 
a classification of global obstructions and their possible
dependence on boundary conditions.

It is interesting to compare this state of affairs with
the gauge-fixing approach.
The latter approach also has succeeded in formulating
abelian ChGTs on the lattice, while it is presently
unknown whether it can be extended to the non-abelian case.
Both methods admit weak-coupling perturbation theory, and
both are perturbatively renormalizable.  In both cases
unitarity is {\it not} manifest, but could in principle
be established perturbatively to all orders.

Of course, in principle, there is a theoretical advantage
in having a manifestly gauge-invariant formulation on the
lattice.  However, even in the abelian case, where progress
has been sufficient to compare the two methods, this
advantage comes at a price.  It is clear that the gauge-invariant 
approach is more difficult to implement numerically, because
it is numerically very expensive to compute the overlap-Dirac
operator, or equivalently, $\ghat$.
Other approaches, such as gauge fixing, may be technically
simpler, even if they lack exact gauge invariance at the
regularized level.  It is important to remember that what is required 
is only the recovery of
gauge invariance in the continuum limit, \ie\ at scales much
below the cutoff $1/a$.  An assumption here is, of course, that
the various different ways of putting ChGTs on the lattice
all describe the same universality class.

At present, there exists two types of gauge non-invariant methods,
the one based on gauge fixing discussed above, 
and the two-cutoff or interpolation
approach.  The way in which the gauge-fixing methods avoids the
Nielsen--Ninomiya theorem is well understood, and it has been
demonstrated both analytically and numerically that chiral fermions
can be put on the lattice.  The main problem lies in the extension
of this approach to non-abelian ChGTs, because of the issue of
Gribov copies.  This cannot be said of the two-cutoff approach,
in which the behavior of the fermion spectrum is not fully
understood beyond perturbation theory.  The work of ref.~\cite{herbou}
has contributed significantly to a clarification of what the
important open problems are in this respect.  I believe that it
is important to gain more analytic insight in this case in order
to assess its viability.

An issue which I did not touch on in this review is that of
fermion-number violation.  In my view, this issue is {\it not}
central to the problem of constructing lattice ChGTs.  If an
approach is successful, this particular problem ``will take care
of itself."  This does not imply that it is not interesting 
to understand in detail how this works out for a given approach.
For general discussions of the problem within the context of
lattice ChGTs, see refs.~\cite{ban,dugman,bhs}.  In the gauge
invariant approach, fermion number is violated just as it is
in the continuum, through the measure of the fermionic path
integral \cite{lueab}.  For the two-cutoff approach, see 
refs.~\cite{hersun,bod}, and for the gauge-fixing approach,
see ref.~\cite{bgspt}.

Finally, there is the question whether it will ever be possible
to investigate ChGTs dynamically on the lattice.  After all,
while it is interesting to learn how ChGTs may be
put on the lattice, one would like to learn more about issues
such as dynamical symmetry breaking in ChGTs, for which also 
analytic methods have contributed relatively little insight.
While full-fledged numerical simulations are clearly not 
within reach yet (the hardest problem being that of a 
complex measure \cite{cha}), it may be possible to probe,
within a given approach, the phase of the fermion determinant,
for instance on an ensemble of quenched configurations.  The
behavior of the fluctuations of the phase of the determinant
should tell us something about how different ChGTs are
dynamically from vector-like gauge theories.  Of course, we
are not there yet, since non-abelian ChGTs
are more interesting than abelian ChGTs, and
the task of putting non-abelian ChGTs 
on the lattice has not yet been completed. 
 
\medskip
\leftline{\it Acknowledgements}
I thank the organizers of Lattice 2000 for giving me the 
opportunity to present this review.  I also thank Oliver B\"ar,
Subhasis Basak, Geoff Bodwin, Michael Creutz, Asit De, 
Robert Edwards, Pilar Hern\'andez, Ji\v r\'\i\ Jers\'ak, Yoshio 
Kikukawa, Martin L\"uscher, Jun Nishimura, Michael Ogilvie,
Noam Shoresh, Hiroshi Suzuki, and, in particular Yigal Shamir 
and Steve Sharpe, for discussions and comments,  as well as 
the INT at the Univ. of Washington for hospitality. This work 
is supported in part by the US Dept. of Energy.

\end{document}